\documentclass{article}%
\usepackage{amsmath}
\usepackage{amsfonts}
\usepackage{amssymb}
\usepackage{graphicx}%
\providecommand{\U}[1]{\protect\rule{.1in}{.1in}}
\newtheorem{theorem}{Theorem}

\newtheorem{remark}[theorem]{Remark}

\begin{document}

\title{Equations of Motion and Energy-Momentum 1-Forms for the Coupled Gravitational,
Maxwell and Dirac Fields}
\author{Waldyr A. Rodrigues Jr. and Samuel A. Wainer\\Institute of Mathematics Statistics and Scientific Computation\\IMECC-UNICAMP\\e-mail: walrod@ime.unicamp.br~~~~samuelwainer@ime.unicamp.br}
\date{January 21 2016}
\maketitle

\begin{abstract}
A theory where the gravitational, Maxwell and Dirac fields (mathematically
represented as particular sections of a convenient Clifford bundle) are
supposed fields in Faraday's sense living in Minkowski spacetime is presented.
In our theory there exist a genuine energy-momentum tensor for the
gravitational field and a genuine energy-momentum conservation law for the
system of the interacting gravitational, Maxwell and Dirac fields. Moreover,
the energy-mometum tensors of the Maxwell and Dirac fields are symmetric, and
it is shown that the equations of motion for the gravitational potentials is
equivalent to Einstein equation of General Relativity (where the second member
is the sum of the energy-momentum tensors of the Maxwell, Dirac and
interaction Maxwell-Dirac fields) defined in an effective Lorentzian
spacetime, whose use\ is eventually no more than a question of mathematical convenience.

\end{abstract}

\section{Introduction}

In this paper we present a theory where the gravitational, Maxwell and Dirac
fields are intepreted as fields in the Faraday sense living and interacting in
Minkowski spacetime structure $(M,\boldsymbol{\mathring{g}},\mathring{D}%
,\tau_{\boldsymbol{\mathring{g},}}\uparrow_{e_{0}})$ (see Appendix A). The
Lagrangian density\footnote{Natural units are used in this paper.} of these
fields are postulated and their energy-momentum tensors are evaluated. All
fields in our theory are mathematically described by sections of a particular
and convenient Clifford bundle $\mathcal{C\ell}(M,\mathtt{g})$ (see Appendix
A) which is used as a mathematical \emph{tool}. In particular the
gravitational field is represented by its gravitational potentials
$\mathfrak{g}^{\mathbf{a}}$, $\mathbf{a}=0,1,2,3$. It is very important to
emphasize here that in our theory we have a genuine energy-mometum
conservation law for the interacting system of the gravitational, Maxwell and
Dirac fields. Moreover, the energy-momentum tensor of the Dirac field in the
presence of the gravitational field is symmetric. It is also very important to
emphasize that the formulation of our theory \emph{does not use} at any time
any connection defined in $M$. However, we may interpret the structure
$(M,\boldsymbol{g},D,\tau_{\boldsymbol{g,}}\uparrow_{\mathfrak{e}_{0}})$
(where $D$ is the Levi-Civita connection of $\boldsymbol{g=}\eta_{\mathbf{ab}%
}\mathfrak{g}^{\mathbf{a}}\otimes\mathfrak{g}^{\mathbf{b}}$, $\tau
_{\boldsymbol{g,}}=\mathfrak{g}^{\mathbf{0}}\mathfrak{g}^{\mathbf{1}%
}\mathfrak{g}^{\mathbf{2}}\mathfrak{g}^{\mathbf{2}}\in\sec%
%TCIMACRO{\tbigwedge ^{r}}%
%BeginExpansion
{\textstyle\bigwedge^{r}}
%EndExpansion
T^{\ast}M\hookrightarrow\sec\mathcal{C\ell(}M,\mathtt{g})$ defines a positive
orientation for $M$ and $\uparrow_{\mathfrak{e}_{0}}$defines a time
orientation, given by the global vector field $\mathfrak{e}_{0}$) as a
\emph{Lorentzian spacetime }representing a gravitational field generated by
the matter energy-momentum tensor as in General Relativity theory\emph{.}This
statement is proved by showing (see details, e.g., in \cite{rodcap} ) that the
equation for the gravitational potentials $\mathfrak{g}^{\mathbf{a}}$
generated by the energy-momentum tensor of the Dirac and Maxwell fields (and
their interaction) is equivalent to Einstein equation in General Relativity
theory. This result is particularly since it permit us to conclude that the
energy-momentum tensor of the Dirac field in our theory is symmetrical (see
Appendix B). Also, with the introduction of the structure $(M,\boldsymbol{g}%
,D,\tau_{\boldsymbol{g,}}\uparrow_{\mathfrak{e}_{0}})$ in our theory it is
possible to encode the energy-mometum 1-form fields for the gravitational
field coming from the awful Eqs.(\ref{10.13}) and \ref{2.7} in a simple and
nice formula as given by Eq.(\ref{2.14}). The paper has three sections and
three appendices. In Section 1 we present the Lagrangian densities for the
coupled gravitational, Maxwell and Dirac fields. In Section 2 we present the
energy-momentum$1$-forms for the gravitational, Maxwell and Dirac fields and
the energy-mometum 1-forms for the interaction between the Maxwell and Dirac
field. \ Section 3 present our conclusions. Appendix A presents the notations
we used and recall some results important for the intelligibility of the
paper. As already said above the detailed evaluation of the energy-momentum
1-forms for the Dirac field is given in Appendix B. Finally in Appendix C we
use the nice formula Eq.(\ref{2.14}) to evaluate the energy of the
Schwarzschild gravitational field for a star of mas $M$ and radius greater
than its Schwarzschild radius.

\section{Lagrangian Densities and Equations of Motion for the Coupled
Gravitational, Maxwell and Dirac Fields}

The Lagrangian density for the coupled gravitational, Dirac and Maxwell fields is:%

\begin{equation}
\mathfrak{L=\mathfrak{L}_{g}+\mathfrak{L}}_{M}\mathfrak{+\mathfrak{L}}%
_{D}+\mathfrak{L}_{FD}=\mathfrak{\mathfrak{L}_{g}+\mathfrak{L}}_{m}%
.\label{2.1}%
\end{equation}
With $\mathfrak{g}^{\mathbf{a}}\in\sec%
%TCIMACRO{\tbigwedge \nolimits^{1}}%
%BeginExpansion
{\textstyle\bigwedge\nolimits^{1}}
%EndExpansion
T^{\ast}M\hookrightarrow\sec\mathcal{C\ell}(M,\mathtt{g}),$ $\mathbf{a}%
=0,1,2,3$ we have\footnote{See Appendix A for notations used in this paper.}%

\begin{gather}
\mathfrak{L}_{\boldsymbol{g}}:(\mathfrak{g}^{\mathbf{a}},d\mathfrak{g}%
^{\mathbf{a}})\mapsto\mathfrak{L}_{\boldsymbol{g}}(\mathfrak{g}^{\mathbf{a}%
},d\mathfrak{g}^{\mathbf{a}})\in\sec%
%TCIMACRO{\tbigwedge \nolimits^{4}}%
%BeginExpansion
{\textstyle\bigwedge\nolimits^{4}}
%EndExpansion
T^{\ast}M.\nonumber\\
\mathfrak{L}_{\boldsymbol{g}}(\mathfrak{g}^{\mathbf{a}},d\mathfrak{g}%
^{\mathbf{a}})=-\frac{1}{2}d\mathfrak{g}^{\mathbf{a}}\wedge
\underset{\boldsymbol{g}}{\star}d\mathfrak{g}_{\mathbf{a}}+\frac{1}%
{2}\underset{\boldsymbol{g}}{\delta}\mathfrak{g}^{\mathbf{a}}\wedge
\underset{\boldsymbol{g}}{\star}\underset{\boldsymbol{g}}{\delta}%
\mathfrak{g}_{\mathbf{a}}+\frac{1}{4}\left(  d\mathfrak{g}^{\mathbf{a}}%
\wedge\mathfrak{g}_{\mathbf{a}}\right)  \wedge\underset{\boldsymbol{g}}{\star
}\left(  d\mathfrak{g}^{\mathbf{b}}\wedge\mathfrak{g}_{\mathbf{b}}\right)
,\label{2.2}%
\end{gather}
Also, with $F\in\sec%
%TCIMACRO{\tbigwedge \nolimits^{2}}%
%BeginExpansion
{\textstyle\bigwedge\nolimits^{2}}
%EndExpansion
T^{\ast}M\hookrightarrow\sec\mathcal{C\ell}(M,\mathtt{g})$%
\begin{gather}
\mathfrak{L}_{M}:F\mapsto\mathfrak{L}_{F}(F)\in\sec%
%TCIMACRO{\tbigwedge \nolimits^{4}}%
%BeginExpansion
{\textstyle\bigwedge\nolimits^{4}}
%EndExpansion
T^{\ast}M,\nonumber\\
\mathfrak{L}_{M}(F)=-\frac{1}{2}F\wedge\underset{\boldsymbol{g}}{\star
}F,\label{2.3}%
\end{gather}
and with $\psi\in\mathcal{C\ell}^{0}(M,\mathtt{g})$ a \emph{representative} in
the Clifford bundle of a Dirac-Hestenes spinor field (once a spin frame is fixed)%

\begin{gather}
(\mathfrak{g}^{\mathbf{k}},\psi,\tilde{\psi},\mathfrak{g}^{\mathbf{k}}%
\partial_{\mathfrak{e}_{\mathbf{k}}}\psi,\mathfrak{g}^{\mathbf{k}}%
\partial_{\mathfrak{e}_{\mathbf{k}}}\tilde{\psi})\mapsto\mathcal{L}%
_{D}(\mathfrak{g}^{\mathbf{k}},\psi,\tilde{\psi},\mathfrak{g}^{\mathbf{k}%
}\partial_{\mathfrak{e}_{\mathbf{k}}}\psi,\mathfrak{g}^{\mathbf{k}}%
\partial_{\mathfrak{e}_{\mathbf{k}}}\tilde{\psi})\in\sec%
%TCIMACRO{\tbigwedge \nolimits^{4}}%
%BeginExpansion
{\textstyle\bigwedge\nolimits^{4}}
%EndExpansion
T^{\ast}M,\nonumber\\
\mathcal{L}_{D}(\mathfrak{g}^{\mathbf{k}},\psi,\tilde{\psi},\mathfrak{g}%
^{\mathbf{k}}\partial_{\mathfrak{e}_{\mathbf{k}}}\psi,\mathfrak{g}%
^{\mathbf{k}}\partial_{\mathfrak{e}_{\mathbf{k}}}\tilde{\psi})\nonumber\\
=\frac{1}{2}\left\{
\begin{array}
[c]{c}%
(\mathfrak{g}^{\mathbf{k}}\partial_{\mathfrak{e}_{\mathbf{k}}}\tilde{\psi
}\mathfrak{g}^{\mathbf{2}}\mathfrak{g}^{\mathbf{1}})\mathfrak{g}^{\mathbf{0}%
}\cdot\tilde{\psi}-\frac{1}{4}\mathfrak{g}^{\mathbf{k}}\tilde{\psi
}L(\mathfrak{g}_{\mathbf{k}})\mathfrak{g}^{\mathbf{0}}\mathfrak{g}%
^{\mathbf{2}}\mathfrak{g}^{\mathbf{1}}\cdot\tilde{\psi}\\
+\psi\cdot(\mathfrak{g}^{\mathbf{k}}\partial_{\mathfrak{e}_{\mathbf{k}}}%
\psi\mathfrak{g}^{\mathbf{0}}\mathfrak{g}^{\mathbf{2}}\mathfrak{g}%
^{\mathbf{1}})+\frac{1}{4}\psi\cdot(\mathfrak{g}^{\mathbf{k}}L(\mathfrak{g}%
_{\mathbf{k}})\psi\mathfrak{g}^{\mathbf{0}}\mathfrak{g}^{\mathbf{2}%
}\mathfrak{g}^{\mathbf{1}}+m\psi\cdot\tilde{\psi}%
\end{array}
\right\}  \tau_{\boldsymbol{g}}\label{2.4A}%
\end{gather}
where the symbol $L(\mathfrak{g}_{\mathbf{k}})$ is defined in the Appendix B
(see Eqs.(\ref{A5}) and (\ref{A6})) and $m$ is the mass of the fermion field.

The interaction Lagrangian density between the Dirac and Maxwell field is
\begin{gather}
\mathfrak{L}_{FD}:(\psi,\tilde{\psi},\mathfrak{g}^{\mathbf{0}},A)\mapsto
\mathfrak{L}_{FD}(\psi,\tilde{\psi},\mathfrak{g}^{\mathbf{0}},A)\in\sec%
%TCIMACRO{\tbigwedge \nolimits^{4}}%
%BeginExpansion
{\textstyle\bigwedge\nolimits^{4}}
%EndExpansion
T^{\ast}M.\nonumber\\
\mathfrak{L}_{FD}(\psi,\tilde{\psi},\mathfrak{g}^{\mathbf{0}},A)=e\tilde{\psi
}\mathfrak{g}^{\mathbf{0}}\psi\wedge\underset{\boldsymbol{g}}{\star
}A\label{2.5}%
\end{gather}
where $e$ is the charge of the fermion field and $A\in\sec%
%TCIMACRO{\tbigwedge \nolimits^{1}}%
%BeginExpansion
{\textstyle\bigwedge\nolimits^{1}}
%EndExpansion
T^{\ast}M\hookrightarrow\sec\mathcal{C\ell}(M,\mathtt{g})$ is the
electromagnetic potential such that $F:=dA\in\sec%
%TCIMACRO{\tbigwedge \nolimits^{2}}%
%BeginExpansion
{\textstyle\bigwedge\nolimits^{2}}
%EndExpansion
T^{\ast}M\hookrightarrow\sec\mathcal{C\ell}(M,\mathtt{g})$.

In our theory it is supposed that at least one of $\mathfrak{g}^{\mathbf{a}}$
is not closed, i.e., $d\mathfrak{g}^{\mathbf{a}}\neq0$, for some
$\mathbf{a}=0,1,2,3$. Putting $\mathcal{F}^{\mathbf{d}}=d\mathfrak{g}%
^{\mathbf{d}}.$the equation of motion for the gravitational potentials are
obtained from the variational principle. We have
\begin{equation}
\boldsymbol{\delta}\int\mathfrak{L}_{\boldsymbol{g}}=\int\boldsymbol{\delta
}\mathfrak{L}_{\boldsymbol{g}}=\int\boldsymbol{\delta}\mathfrak{g}%
^{\mathbf{d}}\wedge\left(  \frac{\boldsymbol{\delta}\mathfrak{L}%
_{\boldsymbol{g}}}{\boldsymbol{\delta}\mathfrak{g}^{\mathbf{d}}}%
+\frac{\boldsymbol{\delta}\mathfrak{L}_{m}}{\boldsymbol{\delta}\mathfrak{g}%
^{\mathbf{d}}}\right)  ,\label{2.v1}%
\end{equation}
where%

\begin{equation}
\underset{\boldsymbol{g}}{\star}%
%TCIMACRO{\tsum \nolimits_{\mathbf{d}}}%
%BeginExpansion
{\textstyle\sum\nolimits_{\mathbf{d}}}
%EndExpansion
=\frac{\boldsymbol{\delta}\mathfrak{L}_{\boldsymbol{g}}}{\boldsymbol{\delta
}\mathfrak{g}^{\mathbf{d}}}=-\left(  \frac{\partial\mathfrak{L}%
_{\boldsymbol{g}}}{\partial\mathfrak{g}^{\mathbf{d}}}+d\left(  \frac
{\partial\mathfrak{L}_{\boldsymbol{g}}}{\partial d\mathfrak{g}^{\mathbf{d}}%
}\right)  \right)  \label{2.v2}%
\end{equation}
is the Euler-Lagrange functional and\footnote{We suppose that $\mathcal{L}%
_{m}$ does not depend explicitly on the $d\mathfrak{g}^{\mathbf{a}}$.}
\begin{align}
\underset{\boldsymbol{g}}{\star}\overset{m}{T}_{\mathbf{d}}  &
=-\underset{\boldsymbol{g}}{\star}\overset{m}{\mathcal{T}}_{\mathbf{d}}%
=\frac{\partial\mathcal{L}_{\boldsymbol{m}}}{\partial\mathfrak{g}^{\mathbf{d}%
}}=\underset{\boldsymbol{g}}{\star}\overset{D}{\mathcal{T}}_{\mathbf{d}%
}+\underset{\boldsymbol{g}}{\star}\overset{M}{\mathcal{T}}_{\mathbf{d}%
}+\underset{\boldsymbol{g}}{\star}\overset{MD}{\mathcal{T}}_{\mathbf{d}%
}\label{2.v3}\\
&  =\frac{\partial\mathcal{L}_{D}}{\partial\mathfrak{g}^{\mathbf{d}}}%
+\frac{\partial\mathcal{L}_{M}}{\partial\mathfrak{g}^{\mathbf{d}}}%
+\frac{\partial\mathcal{L}_{MD}}{\partial\mathfrak{g}^{\mathbf{d}}}%
\end{align}
will be called the energy momentum $3$-forms of the matter fields of the
matter fields. One can show\footnote{Details may be found, e.g., in
\cite{rodcap}} that the equations of motion for the gravitational potentials
coming from $\underset{\boldsymbol{g}}{\star}%
%TCIMACRO{\tsum \nolimits_{\mathbf{d}}}%
%BeginExpansion
{\textstyle\sum\nolimits_{\mathbf{d}}}
%EndExpansion
=0$ are:%

\begin{equation}
-d\underset{\boldsymbol{g}}{\star}\mathcal{S}_{\mathbf{d}}%
-\underset{\boldsymbol{g}}{\star}t_{\mathbf{d}}=\underset{\boldsymbol{g}%
}{\star}\mathcal{T}_{\mathbf{d}}=-\underset{\boldsymbol{g}}{\star
}T_{\mathbf{d}}, \label{10.12}%
\end{equation}
with%
\begin{gather}
\underset{\boldsymbol{g}}{\star}t_{\mathbf{d}}:=\frac{\partial\mathcal{L}%
_{\boldsymbol{g}}}{\partial\mathfrak{g}^{\mathbf{d}}}=\frac{1}{2}%
[(\mathfrak{g}_{\mathbf{d}}\underset{\boldsymbol{g}}{\lrcorner}d\mathfrak{g}%
^{\mathbf{a}})\wedge\underset{\boldsymbol{g}}{\star}d\mathfrak{g}_{\mathbf{a}%
}-d\mathfrak{g}^{\mathbf{a}}\wedge(\mathfrak{g}_{\mathbf{d}}%
\underset{\boldsymbol{g}}{\lrcorner}\underset{\boldsymbol{g}}{\star
}d\mathfrak{g}_{\mathbf{a}})]\nonumber\\
+\frac{1}{2}d(\mathfrak{g}_{\mathbf{d}}\underset{\boldsymbol{g}}{\lrcorner
}\underset{\boldsymbol{g}}{\star}\mathfrak{g}^{\mathbf{a}})\wedge
\underset{\boldsymbol{g}}{\star}d\underset{\boldsymbol{g}}{\star}%
\mathfrak{g}_{\mathbf{a}}+\frac{1}{2}(\mathfrak{g}_{\mathbf{d}}%
\underset{\boldsymbol{g}}{\lrcorner}\underset{\boldsymbol{g}}{\star
}\mathfrak{g}^{\mathbf{a}})\wedge\underset{\boldsymbol{g}}{\star
}d\underset{\boldsymbol{g}}{\star}\mathfrak{g}_{\mathbf{a}}+\frac{1}%
{2}d\mathfrak{g}_{\mathbf{d}}\wedge\underset{\boldsymbol{g}}{\star}\left(
d\mathfrak{g}^{\mathbf{a}}\wedge\mathfrak{g}_{\mathbf{a}}\right) \nonumber\\
-\frac{1}{4}d\mathfrak{g}^{\mathbf{a}}\wedge\mathfrak{g}_{\mathbf{a}}%
\wedge\left[  \mathfrak{g}_{\mathbf{d}}\underset{\boldsymbol{g}}{\lrcorner
}\underset{\boldsymbol{g}}{\star}\left(  d\mathfrak{g}^{\mathbf{c}}%
\wedge\mathfrak{g}_{\mathbf{c}}\right)  \right]  -\frac{1}{4}\left[
\mathfrak{g}_{\mathbf{d}}\underset{\boldsymbol{g}}{\lrcorner}\left(
d\mathfrak{g}^{\mathbf{c}}\wedge\mathfrak{g}_{\mathbf{c}}\right)  \right]
\wedge\underset{\boldsymbol{g}}{\star}\left(  d\mathfrak{g}^{\mathbf{a}}%
\wedge\mathfrak{g}_{\mathbf{a}}\right)  , \label{10.13}%
\end{gather}%
\begin{equation}
\underset{\boldsymbol{g}}{\star}\mathcal{S}_{\mathbf{d}}:=\frac{\partial
\mathcal{L}_{\boldsymbol{g}}}{\partial d\mathfrak{g}^{\mathbf{d}}%
}=-\underset{\boldsymbol{g}}{\star}d\mathfrak{g}_{\mathbf{d}}-(\mathfrak{g}%
_{\mathbf{d}}\underset{\boldsymbol{g}}{\lrcorner}\underset{\boldsymbol{g}%
}{\star}\mathfrak{g}^{\mathbf{a}})\wedge\underset{\boldsymbol{g}}{\star
}d\underset{\boldsymbol{g}}{\star}\mathfrak{g}_{\mathbf{a}}+\frac{1}%
{2}\mathfrak{g}_{\mathbf{d}}\wedge\underset{\boldsymbol{g}}{\star}\left(
d\mathfrak{g}^{\mathbf{a}}\wedge\mathfrak{g}_{\mathbf{a}}\right)  .
\label{10.14}%
\end{equation}

Moreover, putting $\mathcal{F}^{\mathbf{a}}:=d\mathfrak{g}^{a}$, it is of
course, $d\mathcal{F}^{\mathbf{a}}=0$ and the field equations (Eq.(\ref{10.12}%
)) can be written as
\begin{equation}
d\underset{\boldsymbol{g}}{\star}\mathcal{F}_{\mathbf{d}}%
=-\underset{\boldsymbol{g}}{\star}\overset{m}{\mathcal{T}}_{\mathbf{d}%
}-\underset{\boldsymbol{g}}{\star}t_{\mathbf{d}}-\underset{\boldsymbol{g}%
}{\star}\mathfrak{h}_{d} \label{10.16}%
\end{equation}
where%
\begin{equation}
\mathfrak{h}_{\mathbf{d}}=d\left[  (\mathfrak{g}_{\mathbf{d}}%
\underset{\boldsymbol{g}}{\lrcorner}\underset{\boldsymbol{g}}{\star
}\mathfrak{g}^{\mathbf{a}})\wedge\underset{\boldsymbol{g}}{\star
}d\underset{\boldsymbol{g}}{\star}\mathfrak{g}_{\mathbf{a}}-\frac{1}%
{2}\mathfrak{g}_{\mathbf{d}}\wedge\underset{\boldsymbol{g}}{\star}\left(
\mathcal{F}^{\mathbf{a}}\wedge\mathfrak{g}_{\mathbf{a}}\right)  \right]  .
\label{10.17}%
\end{equation}
So, we have%

\begin{gather}
(\text{a})~d\mathcal{F}_{\mathbf{d}}=0,~~~~\left(  \text{b}\right)
~\underset{\boldsymbol{g}}{\delta}\mathcal{F}_{\mathbf{d}}=-\left(
\overset{m}{\mathcal{T}}_{\mathbf{d}}+\mathbf{t}_{\mathbf{d}}\right)
,\label{2.6}\\
\mathbf{t}_{\mathbf{d}}=(t_{\mathbf{d}}+\mathfrak{h}_{\mathbf{d}}).
\label{2.7}%
\end{gather}

Also, it is very much important to recall that introducing the Levi-Civita
connection of $\boldsymbol{g=}\eta_{\mathbf{ab}}\mathfrak{g}^{\mathbf{a}%
}\otimes\mathfrak{g}^{\mathbf{b}}$ into the game one can show with some
algebra (details, e.g., in \cite{rodcap}) that
\begin{equation}
-d\underset{\boldsymbol{g}}{\star}\mathcal{S}_{\mathbf{d}}-\text{
}\underset{\boldsymbol{g}}{\star}t_{\mathbf{d}}=\underset{\boldsymbol{g}%
}{\star}G_{\mathbf{d}}\label{2.11}%
\end{equation}
where $G_{\mathbf{d}}:=G_{\mathbf{dk}}\mathfrak{g}^{\mathbf{k}}\in\sec%
%TCIMACRO{\tbigwedge \nolimits^{1}}%
%BeginExpansion
{\textstyle\bigwedge\nolimits^{1}}
%EndExpansion
T^{\ast}M\hookrightarrow\sec\mathcal{C\ell}(M,\mathtt{g})$ are the Einstein
$1$-form fields, with
\begin{equation}
G_{\mathbf{dk}}=R_{\mathbf{dk}}-\frac{1}{2}\eta_{\mathbf{dk}}R=G_{\mathbf{dk}%
}\label{2.12}%
\end{equation}
where $R_{\mathbf{dk}}$ are the components of the Ricci tensor and $R$ is the
scalar curvature in the structure $(M,\boldsymbol{g},D,\tau_{\boldsymbol{g,}%
}\uparrow_{\mathfrak{e}_{0}})$.

With this result we immediately infer from Eq.(\ref{10.12}) that writing
$\overset{m}{\mathcal{T}}_{\mathbf{d}}=\overset{m}{\mathcal{T}}_{\mathbf{dk}%
}\mathfrak{g}^{\mathbf{k}}\in\sec%
%TCIMACRO{\tbigwedge \nolimits^{1}}%
%BeginExpansion
{\textstyle\bigwedge\nolimits^{1}}
%EndExpansion
T^{\ast}M\hookrightarrow\sec\mathcal{C\ell}(M,\mathtt{g})$ it is%
\begin{equation}
\overset{m}{\mathcal{T}}_{\mathbf{dk}}=\overset{m}{\mathcal{T}}_{\mathbf{kd}}.
\label{2.13}%
\end{equation}
an of course we must also have:%
\begin{equation}
\overset{M}{\mathcal{T}}_{\mathbf{dk}}=\overset{M}{\mathcal{T}}_{\mathbf{kd}%
},~~\overset{D}{\mathcal{T}}_{\mathbf{dk}}=\overset{D}{\mathcal{T}%
}_{\mathbf{kd}},~~,\overset{MD}{\mathcal{T}}_{\mathbf{dk}}%
=\overset{MD}{\mathcal{T}}_{\mathbf{kd}}. \label{2.13A}%
\end{equation}

However it is not the case that in general $t_{\mathbf{dk}}=t_{\mathbf{kd}}$.
See Section 3.1.

\begin{remark}
It is crucial to emphasize here that the introduction of a Lorentzian
spacetime structure $(M,\boldsymbol{g},D,\tau_{\boldsymbol{g}},\uparrow
_{\mathfrak{e}_{0}})$ to get \emph{Eq.(\ref{2.13A}) is to be viewed as simple
a mathematical aid, no fundamental ontology is given to that Lorentzian
structure \ Indeed, it has been shown in details, e.g., in
\cite{mol,nrr2010,rod2012} \ that our theory of the gravitational field may be
interpreted as generating spacetime structures with general connections where
curvature torsion and non metricity tensors may be non null.}
\end{remark}

Also, we recall that the equations of motion for the Dirac and Maxwell fields
are respectively (see, e.g., \cite{rodcap} for details of the derivation)
\begin{equation}
\mathfrak{g}^{\mathbf{a}}\mathbf{D}_{\mathfrak{e}_{\mathbf{a}}}\psi
\mathfrak{g}^{\mathbf{2}}\mathfrak{g}^{\mathbf{1}}-m\psi\mathfrak{g}%
^{\mathbf{0}}+eA\psi=0\label{2.D}%
\end{equation}
and%
\begin{gather}
dF=0,~~~\underset{\boldsymbol{g}}{\delta}F=-J_{e,}\nonumber\\
J_{e}=e\psi\mathfrak{g}^{\mathbf{0}}\tilde{\psi}.\label{2.M}%
\end{gather}

\section{Energy-Momentum $1$-Forms Fields for the Gravitational,Maxwell and
Dirac Fields}

\subsection{Gravitational Energy-Momentum $1$-Forms}

Despite the very awful formula for $\mathbf{t}_{\mathbf{d}}$ coming from
Eqs.(\ref{10.13}) and (\ref{2.7}) it has been shown in \cite{rod2012} that
\ it can be coded in a nice simple formula once we introduce as an auxiliary
mathematical device the Levi-Civita connection of $\boldsymbol{g}$ and the
Dirac operator $\boldsymbol{\partial=}\mathfrak{g}^{\mathbf{d}}D_{\mathbf{d}}$
acting on sections of $\mathcal{C}\ell(M,\mathtt{g})$. Indeed, it is:%

\begin{equation}
\mathbf{t}^{\mathbf{d}}=\frac{1}{2}R\mathfrak{g}^{\mathbf{d}}%
+\boldsymbol{\partial}\cdot\boldsymbol{\partial~}\mathfrak{g}^{\mathbf{d}%
}+d\underset{\boldsymbol{g}}{\delta}\mathfrak{g}^{\mathbf{d}} \label{2.14}%
\end{equation}
where $\boldsymbol{\partial}\cdot\boldsymbol{\partial}$ is the covariant
D'Alembertian\cite{rodcap}.

\begin{remark}
It is very important to observe that the objects $\mathbf{t}_{\mathbf{da}%
}=\eta_{\mathbf{ac}}\eta_{\mathbf{dl}}\mathbf{t}^{\mathbf{c}}%
\underset{\boldsymbol{g}}{\lrcorner}\mathfrak{g}^{\mathbf{l}}$ are components
of a legitimate gravitational energy-momentum tensor tensor field
$\mathbf{t}=\mathbf{t}_{\mathbf{da}}\mathfrak{g}^{\mathbf{d}}\otimes
\mathfrak{g}^{\mathbf{a}}\in\sec T_{0}^{2}M$. Also it is worth to take into
account that
\begin{equation}
\mathbf{t}^{\mathbf{da}}-\mathbf{t}^{\mathbf{ad}}=(\boldsymbol{\partial}%
\cdot\boldsymbol{\partial~}\mathfrak{g}^{\mathbf{d}})\underset{\boldsymbol{g}%
}{\lrcorner}\mathfrak{g}^{\mathbf{a}}-(\boldsymbol{\partial}\cdot
\boldsymbol{\partial~}\mathfrak{g}^{\mathbf{a}})\underset{\boldsymbol{g}%
}{\lrcorner}\mathfrak{g}^{\mathbf{d}}+(d\underset{\boldsymbol{g}}{\delta
}\mathfrak{g}^{\mathbf{d}})\underset{\boldsymbol{g}}{\lrcorner}\mathfrak{g}%
^{\mathbf{a}}-(d\underset{\boldsymbol{g}}{\delta}\mathfrak{g}^{\mathbf{a}%
})\underset{\boldsymbol{g}}{\lrcorner}\mathfrak{g}^{\mathbf{b}}\label{2.15}%
\end{equation}
i.e., the energy-momentum tensor of the gravitational field is in general not
symmetric. As observed in \emph{\cite{rodcap}} this is important in order to
have a total angular momentum conservation law for the system consisting of
the gravitational plus the matter fields. In \emph{Appendix C} we present
$\mathbf{t}^{\mathbf{da}}$ for the Schwarzschild solution of Einstein equation
in order to show that it is a viable quantity to really describe the energy
momentum tensor of the gravitational field. In that example it is clear that
$\mathbf{t}^{\mathbf{12}}\neq\mathbf{t}^{\mathbf{21}}$.
\end{remark}

\subsection{Maxwell Energy-Momentum $1$-forms}

We recall moreover that the energy-momentum $1$-forms
$\underset{\boldsymbol{g}}{\star}\overset{M}{T}_{\mathbf{a}}(=-$
$\underset{\boldsymbol{g}}{\star}\overset{M}{\mathcal{T}}_{\mathbf{a}})$ for
the Maxwell field is\footnote{See, e.g., Section 9.9 of \cite{rodcap} for
details of the derivation.}%
\begin{gather}
\underset{\boldsymbol{g}}{\star}\overset{M}{T}_{\mathbf{a}}=-\frac
{\partial\mathcal{L}_{M}}{\partial\mathfrak{g}^{\mathbf{a}}}%
=\underset{\boldsymbol{g}}{\star}\left(  \frac{1}{2}\theta_{\mathbf{a}}(F\cdot
F)+(\theta_{\mathbf{a}}\lrcorner F)\lrcorner F\right)  \label{2.16}\\
=\underset{\boldsymbol{g}}{\star}\left(  \frac{1}{2}F\mathfrak{g}_{\mathbf{a}%
}\tilde{F}\right)  =\left(  \frac{1}{2}F\mathfrak{g}_{\mathbf{a}}\tilde
{F}\right)  \underset{\boldsymbol{g}}{\lrcorner}\tau_{\boldsymbol{g}}%
\end{gather}
and writing $\overset{M}{T}_{\mathbf{a}}=T_{\mathbf{ab}}\mathfrak{g}%
^{\mathbf{b}}$ we get%
\begin{equation}
T_{\mathbf{ab}}=T_{\mathbf{a}}\cdot\mathfrak{g}_{\mathbf{b}}=-\eta
^{\mathbf{cl}}F_{\mathbf{ac}}F_{\mathbf{bl}}+\frac{1}{4}F_{\mathbf{cd}%
}F^{\mathbf{cd}}\eta_{\mathbf{ab}}=T_{\mathbf{b}}\cdot\mathfrak{g}%
_{\mathbf{a}}=T_{\mathbf{ba}}\label{2.17}%
\end{equation}

\subsection{Maxwell-Dirac Interaction Energy-Momentum $1$-forms}

The energy-momentum $1$-forms $\overset{MD}{\mathcal{T}}_{\mathbf{a}}$ are
trivially calculated. We have
\begin{gather}
\overset{MD}{\mathcal{T}}_{\mathbf{a}}=\frac{\partial\mathfrak{L}_{MD}%
}{\partial\mathfrak{g}^{\mathbf{a}}}=eA_{\mathbf{a}}\tilde{\psi}%
\mathfrak{g}^{\mathbf{0}}\psi,\nonumber\\
\overset{MD}{\mathcal{T}}_{\mathbf{ab}}=\frac{1}{2}\left(
\overset{MD}{\mathcal{T}}_{\mathbf{a}}\cdot\mathfrak{g}_{\mathbf{b}%
}+\overset{MD}{\mathcal{T}}_{\mathbf{b}}\cdot\mathfrak{g}_{\mathbf{a}}\right)
=\frac{1}{2}e\langle A_{\mathbf{a}}\tilde{\psi}\mathfrak{g}^{\mathbf{0}}%
\psi\mathfrak{g}_{\mathbf{b}}+A_{\mathbf{b}}\tilde{\psi}\mathfrak{g}%
^{\mathbf{0}}\psi\mathfrak{g}_{\mathbf{a}}\rangle_{1}. \label{2.18}%
\end{gather}

\subsection{Dirac Energy-Momentum $1$-forms}

The calculation of the Dirac energy-momentum $1$-forms is trick and is
presented in Appendix B. We found%

\begin{equation}
\overset{D}{\mathcal{T}}_{\mathbf{k}}=\langle\mathbf{D}_{\mathfrak{e}%
_{\mathbf{k}}}\tilde{\psi}\mathfrak{g}^{\mathbf{2}}\mathfrak{g}^{\mathbf{1}%
}\mathfrak{g}^{\mathbf{0}}\tilde{\psi}+\psi\mathbf{D}_{\mathfrak{e}%
_{\mathbf{k}}}\psi\mathfrak{g}^{\mathbf{0}}\mathfrak{g}^{\mathbf{2}%
}\mathfrak{g}^{\mathbf{1}}\rangle_{1} \label{2.19}%
\end{equation}
and
\begin{equation}
\overset{D}{\mathcal{T}}_{\mathbf{mk}}=\frac{1}{2}\langle\tilde{\psi
}\mathfrak{g}_{(\mathbf{m}}\mathbf{D}_{\mathfrak{e}_{\mathbf{k)}}}%
\psi\mathfrak{g}^{\mathbf{2}}\mathfrak{g}^{\mathbf{1}}\mathfrak{g}%
^{\mathbf{0}}-\mathbf{D}_{(\mathfrak{e}_{\mathbf{k}}}\tilde{\psi}%
\mathfrak{g}_{\mathbf{k)}}\mathfrak{g}^{\mathbf{2}}\mathfrak{g}^{\mathbf{1}%
}\mathfrak{g}^{\mathbf{0}}\psi\rangle_{0}. \label{2.20}%
\end{equation}

Also, it is wort to emphasize that in our theory we have a genuine
conservation law for the energy-momentum of the matter plus the gravitational
field. Indeed it follows from Eq.(\ref{2.6}b) that
\begin{equation}
\underset{\boldsymbol{g}}{\delta}\left(  \overset{m}{\mathcal{T}}_{\mathbf{d}%
}+\mathbf{t}_{\mathbf{d}}\right)  =0. \label{2.21}%
\end{equation}

Finally, it is worth to emphasize that since our spacetime manifold is
parallelizable it its possible to defined a legitimate energy-momentum
covector for the matter plus the gravitational field\footnote{See a detailed
discussion about conservation laws and conditions for existence of an
energy-momentum \emph{covector }(not a covector field) in \cite{rodwai2015}.
\par
{}}, namely
\begin{align}
\boldsymbol{P}  &  \boldsymbol{=}P_{\mathbf{d}}\mathfrak{g}^{\mathbf{d}%
},\nonumber\\
P_{\mathbf{d}}  &  =\int\underset{\boldsymbol{g}}{\star}\left(
\overset{m}{\mathcal{T}}_{\mathbf{d}}+\mathbf{t}_{\mathbf{d}}\right)  .
\label{2.22}%
\end{align}

\section{Conclusions}

In his paper we present a coherent relativistic theory of the gravitational,
Maxwell and Dirac fields in interaction. In our theory field equations and the
corresponding energy-momentum tensors of the fields are obtained from the
variational principle through postulated Lagrangian densities for those fields
and their interactions. All fields are intended as fields in Faraday's sense
living in a Minkowski spacetime structure. The energy-mometum tensors for the
Maxwell and Dirac fields are symmetric and it is recalled that the equations
satisfied by the gravitational potentials are equivalent to Einstein equation
of General Relativity in an effective Lorentzian spacetime structure
$(M,\boldsymbol{g},D,\tau_{\boldsymbol{g,}}\uparrow_{\mathfrak{e}_{0}})$ which
differently from the case of General Relativity is not supposed to have any
ontology, it is used in the paper only as a tool to obtain an important
mathematical result need for the construction of the energy-mometum tensor of
the Dirac field and to obtain a short formula (Eq.(\ref{2.14})) for the
energy-momentum of the gravitational field whose derivation from the
gravitational Lagrangian density produces a somewhat awful (but of course,
correct) formula (see Eq.(\ref{2.13}) and Eq.(\ref{2.7})). Moreover, the
viability of our formula for really representing the energy-momentum of the
gravitational field is shown by explicitly evaluating it for the Schwarzschild
field of a star of mass $M$ and radius much greater than its Schwarzschild radius.

\appendix

\section{Notations and Recall of Some Results}

In this paper $M$ designs a $4$-dimensional manifold diffeomorphic to
$\mathbb{R}^{4}$ whose elements are called events. If $\{\mathrm{x}^{\mu
}\},\mu=0,1,2,3$ are global coordinates for $M$, $\{e_{\mu}\},e_{\mu}%
=\frac{\partial}{\partial\mathrm{x}^{\mu}}\in\sec TM$ are global smooth vector
fields and we denote denote by $\{\theta^{\mu}=d\mathrm{x}^{\mu}\}\in\sec%
%TCIMACRO{\tbigwedge \nolimits^{1}}%
%BeginExpansion
{\textstyle\bigwedge\nolimits^{1}}
%EndExpansion
T^{\ast}M$ its dual basis. We can introduce in $M$ several different metric
fields, in particular\ an euclidean metric field
\[
\boldsymbol{\mathring{g}}_{E}=\delta_{\mu\nu}\theta^{\mu}\otimes\theta^{\nu
}\in\sec T_{2}^{0}M
\]
and also a Lorentzian metric field
\begin{equation}
\boldsymbol{\mathring{g}=}\eta_{\mu\nu}\theta^{\mu}\otimes\theta^{\nu}\in\sec
T_{2}^{0}M \label{m0}%
\end{equation}
of signature\footnote{This means that the matrix with entries $\eta_{\mu\nu}$
is the diagonal matrix $(\eta_{\mu\nu})=\mathrm{diag}(1,-1,-1,-1)$. Also if
$\eta^{\mu\nu}\eta_{\nu\alpha}=\delta_{\alpha}^{\mu}$, then the matrix with
entries $\left(  \eta^{\mu\nu}\right)  =\mathrm{diag}(1,-1,-1,-1).$} $-2$.

We denoted by $\mathtt{\mathring{g}}_{E},\mathtt{\mathring{g}}$ $\mathtt{\in
}\sec T_{0}^{2}M$ metrics on the cotangent bundle such that
\begin{equation}
\mathtt{\mathring{g}}_{E}=\delta^{\mu\nu}e_{\mu}\otimes e_{\nu}%
,~~\mathtt{\mathring{g}}=\eta^{\mu\nu}e_{\mu}\otimes e_{\nu}.\label{m02}%
\end{equation}

Moreover, we denote by $\mathring{g}$ the extensor field
\[
\mathring{g}:\sec%
%TCIMACRO{\tbigwedge \nolimits^{1}}%
%BeginExpansion
{\textstyle\bigwedge\nolimits^{1}}
%EndExpansion
T^{\ast}M\rightarrow\sec%
%TCIMACRO{\tbigwedge \nolimits^{1}}%
%BeginExpansion
{\textstyle\bigwedge\nolimits^{1}}
%EndExpansion
T^{\ast}M,
\]
such that for $a,b\in\sec%
%TCIMACRO{\tbigwedge \nolimits^{1}}%
%BeginExpansion
{\textstyle\bigwedge\nolimits^{1}}
%EndExpansion
T^{\ast}M$ it is\footnote{We define for $a,b\in\sec%
%TCIMACRO{\tbigwedge \nolimits^{1}}%
%BeginExpansion
{\textstyle\bigwedge\nolimits^{1}}
%EndExpansion
T^{\ast}M,\mathtt{\mathring{g}}_{E}(a,b):=a\bullet b$.}%
\begin{equation}
\mathring{g}(a)\bullet b:=\mathtt{\mathring{g}}%
(a,b):=a\underset{\boldsymbol{\mathring{g}}}{\bullet}b~.\label{m03}%
\end{equation}
Of course we can introduce in the structure $(M,\boldsymbol{\mathring{g}}%
_{E})$ [respectively $(M,\boldsymbol{\mathring{g}}_{E})$] the Clifford
bundles\footnote{$\mathcal{C\ell(}M,\mathtt{g}_{E})$has been called in
\cite{femoro2001} the canonical agebra.} $\mathcal{C\ell(}M,\mathtt{g}_{E})$
[respectively $\mathcal{C\ell(}M,\mathtt{\mathring{g}})]$ and of course, we
have that $%
%TCIMACRO{\tbigwedge }%
%BeginExpansion
{\textstyle\bigwedge}
%EndExpansion
T^{\ast}M$, the bundle of exterior forms is such that\footnote{Details in
\cite{rodcap}.}) $%
%TCIMACRO{\tbigwedge }%
%BeginExpansion
{\textstyle\bigwedge}
%EndExpansion
T^{\ast}M=%
%TCIMACRO{\tsum \nolimits_{r=0}^{4}}%
%BeginExpansion
{\textstyle\sum\nolimits_{r=0}^{4}}
%EndExpansion%
%TCIMACRO{\tbigwedge ^{r}}%
%BeginExpansion
{\textstyle\bigwedge^{r}}
%EndExpansion
T^{\ast}M\hookrightarrow\mathcal{C\ell(}M,\mathtt{\mathring{g}}_{E})$
[respectively $%
%TCIMACRO{\tbigwedge }%
%BeginExpansion
{\textstyle\bigwedge}
%EndExpansion
T^{\ast}M=%
%TCIMACRO{\tsum \nolimits_{r=0}^{4}}%
%BeginExpansion
{\textstyle\sum\nolimits_{r=0}^{4}}
%EndExpansion%
%TCIMACRO{\tbigwedge ^{r}}%
%BeginExpansion
{\textstyle\bigwedge^{r}}
%EndExpansion
T^{\ast}M\hookrightarrow\mathcal{C\ell(}M,\mathtt{\mathring{g}})$]

Following the ideas presented in \cite{fero} the gravitational field generated
by an energy-momentum tensor $\boldsymbol{T}\in\sec T_{2}^{0}M$ is represented
by a gauge extensor (deformation extensor)\footnote{The $\boldsymbol{h}$
extensor field produces a \emph{plastic} \emph{distortion} of the Lorentz
vacuum (which is defined as the Minkowski spacetime structure). Details in
\cite{fero}.}
\begin{equation}
\boldsymbol{h}:\sec%
%TCIMACRO{\tbigwedge \nolimits^{1}}%
%BeginExpansion
{\textstyle\bigwedge\nolimits^{1}}
%EndExpansion
T^{\ast}M\rightarrow\sec%
%TCIMACRO{\tbigwedge \nolimits^{1}}%
%BeginExpansion
{\textstyle\bigwedge\nolimits^{1}}
%EndExpansion
T^{\ast}M\label{m1}%
\end{equation}
such that putting $\theta^{\mathbf{a}}:=\delta_{\mu}^{\mathbf{a}}%
d\mathrm{x}^{\mu},(\mathbf{a}=0,1,2,3)$ it is.
\begin{equation}
\boldsymbol{h}(\theta^{\mathbf{a}})=\mathfrak{g}^{\mathbf{a}}.\label{m2}%
\end{equation}
The set $\{\mathfrak{g}^{\mathbf{a}}\}$ are called gravitational potentials.
We introduce in $M$ the field $\boldsymbol{g}\in\sec T_{2}^{0}M$ according to
the definition%
\begin{equation}
\boldsymbol{g=}\eta_{\mathbf{ab}}\mathfrak{g}^{\mathbf{a}}\otimes
\mathfrak{g}^{\mathbf{b}}.\label{m3}%
\end{equation}

If $\{\mathfrak{e}_{\mathbf{a}}\}\in\sec TM$ is the dual basis of
$\{\mathfrak{g}^{\mathbf{a}}\}$ we define a field $\mathtt{g}\in\sec T_{0}%
^{2}M$ such that
\begin{equation}
\mathtt{g}=\eta^{a\mathbf{b}}\mathfrak{e}_{\mathbf{a}}\otimes\mathfrak{e}_{b}.
\label{m4}%
\end{equation}

\subsubsection{The Clifford Bundle of Differential forms $\mathcal{C\ell
(}M,\mathtt{g})$}

Since of course, the structure $\mathcal{(}M,\mathtt{g})$ is parallelizable we
can present the Clifford bundle of differential forms as the vector
bundle\footnote{A general section of $\mathcal{C\ell(}M,\mathtt{g})$ \ is a
sum of nonhomogeous differential forms, called multiform fields or Clifford
fields.} $\mathcal{C\ell(}M,\mathtt{g})=P_{\mathrm{Spin}_{1,3}^{e}%
}(M,\boldsymbol{g)}\times_{\mathrm{Ad}^{\prime}}\mathbb{R}_{1,3}$, where
$P_{\mathrm{Spin}_{1,3}^{e}}(M,\boldsymbol{g)}$ is the spin structure bundle
and $\mathbb{R}_{1,3}\simeq\mathbb{H(}2)$ is the so called spacetime algebra.
We recall that \ we also have that $%
%TCIMACRO{\tbigwedge }%
%BeginExpansion
{\textstyle\bigwedge}
%EndExpansion
T^{\ast}M=%
%TCIMACRO{\tsum \nolimits_{r=0}^{4}}%
%BeginExpansion
{\textstyle\sum\nolimits_{r=0}^{4}}
%EndExpansion%
%TCIMACRO{\tbigwedge ^{r}}%
%BeginExpansion
{\textstyle\bigwedge^{r}}
%EndExpansion
T^{\ast}M\hookrightarrow\mathcal{C\ell(}M,\mathtt{g})$.

Given the structure $(M,\boldsymbol{\mathring{g}}_{E})$ with $%
%TCIMACRO{\tbigwedge }%
%BeginExpansion
{\textstyle\bigwedge}
%EndExpansion
T^{\ast}M\hookrightarrow\mathcal{C\ell(}M,\mathtt{\mathring{g}})$ we denoted
by $g:=\boldsymbol{h}^{\dagger}\boldsymbol{\mathring{g}}_{E}\boldsymbol{h}$
the extensor field
\begin{equation}
g:\sec%
%TCIMACRO{\tbigwedge \nolimits^{1}}%
%BeginExpansion
{\textstyle\bigwedge\nolimits^{1}}
%EndExpansion
T^{\ast}M\rightarrow\sec%
%TCIMACRO{\tbigwedge \nolimits^{1}}%
%BeginExpansion
{\textstyle\bigwedge\nolimits^{1}}
%EndExpansion
T^{\ast}M \label{M5}%
\end{equation}
such that for $a,b\in\sec%
%TCIMACRO{\tbigwedge \nolimits^{1}}%
%BeginExpansion
{\textstyle\bigwedge\nolimits^{1}}
%EndExpansion
T^{\ast}M$ it is%
\begin{equation}
\mathtt{g}(a,b):=g(a)\bullet b=\boldsymbol{h}^{\dagger}\boldsymbol{\mathring
{g}h}(a)\bullet(a)=\boldsymbol{\mathring{g}h}(a)\bullet\boldsymbol{h}%
(a)\nonumber
\end{equation}

Also, given the structure $(M,\boldsymbol{\mathring{g}})$ with $%
%TCIMACRO{\tbigwedge }%
%BeginExpansion
{\textstyle\bigwedge}
%EndExpansion
T^{\ast}M\hookrightarrow\mathcal{C\ell(}M,\mathtt{\mathring{g}})$ we \ may
denote by $g:=\boldsymbol{h}^{\dagger}\boldsymbol{h}$ the extensor field
\[
g:\sec%
%TCIMACRO{\tbigwedge \nolimits^{1}}%
%BeginExpansion
{\textstyle\bigwedge\nolimits^{1}}
%EndExpansion
T^{\ast}M\rightarrow\sec%
%TCIMACRO{\tbigwedge \nolimits^{1}}%
%BeginExpansion
{\textstyle\bigwedge\nolimits^{1}}
%EndExpansion
T^{\ast}M
\]
such that for $a,b\in\sec%
%TCIMACRO{\tbigwedge \nolimits^{1}}%
%BeginExpansion
{\textstyle\bigwedge\nolimits^{1}}
%EndExpansion
T^{\ast}M$ it is%
\begin{equation}
g(a)\underset{\boldsymbol{\mathring{g}}}{\bullet}b=\boldsymbol{h}%
(a)\underset{\boldsymbol{\mathring{g}}}{\bullet}\boldsymbol{h}(a)=\mathtt{g}%
(a,b):=a\cdot b. \label{m5}%
\end{equation}

The above relations are essential for the formalism used in \cite{fero} where
a Lagrangian formalism for the $\boldsymbol{h}$ field is developed.
Unfortunately to grasp that theory it is first necessary to have a working
knowledge of the (nontrivial) mathematical theory of extensor fields and
extensor functionals. So in this paper we present the gravitational theory
formulated through the gravitational potentials $\mathfrak{g}^{\mathbf{a}}$
(which is a relatively simple theory) for which the Lagrangian density given
by Eq.(\ref{2.2}) is postulated.\medskip

With $a,b\in\sec%
%TCIMACRO{\tbigwedge \nolimits^{1}}%
%BeginExpansion
{\textstyle\bigwedge\nolimits^{1}}
%EndExpansion
T^{\ast}M\hookrightarrow\mathcal{C\ell(}M,\mathtt{g})$ we have the fundamental
relation\footnote{In this paper the Clifford product is denoted by
juxtaposition of symbols. A detailed explanation of all symbols and identities
need for the derivatins in this paper can be found in \cite{rodcap}}%
\begin{equation}
ab+ba=2\mathtt{g}(a,b)\label{mc1}%
\end{equation}
and moreover%
\begin{equation}
a\cdot b=\frac{1}{2}(ab+ba),~~~a\wedge b=\frac{1}{2}(ab-ba).\label{mc2}%
\end{equation}

A general section of $\mathcal{C\ell(}M,\mathtt{g})$ is written as a sum of
nonhomogeneous differential forms, i.e.,%
\begin{align}
\mathcal{C} &  \mathcal{=}%
%TCIMACRO{\tsum \nolimits_{\mathbf{J}}}%
%BeginExpansion
{\textstyle\sum\nolimits_{\mathbf{J}}}
%EndExpansion
\mathcal{C}_{\mathbf{j}}\mathfrak{g}^{\mathbf{J}}=%
%TCIMACRO{\tsum \nolimits_{\mathbf{J}}}%
%BeginExpansion
{\textstyle\sum\nolimits_{\mathbf{J}}}
%EndExpansion
\mathcal{C}^{\mathbf{J}}\mathfrak{g}_{\mathbf{J}},\nonumber\\
\mathcal{C}_{\mathbf{j}},\mathcal{C}^{\mathbf{J}} &  \in\sec%
%TCIMACRO{\tbigwedge \nolimits^{0}}%
%BeginExpansion
{\textstyle\bigwedge\nolimits^{0}}
%EndExpansion
T^{\ast}M\hookrightarrow\sec\mathcal{C\ell(}M,\mathtt{g})\label{a2A}%
\end{align}
where the symbol $\mathbf{J}$ denotes collective indices. Recall, e.g.,
that\footnote{The concept of the Lie derivative of spinor fields is a subtle
one, with many non equivalent definitions. See a sample of the bibliography in
\cite{lrw2015}. In particular it is even possible \cite{daper} to give a
meaning to a statement one find in physical textbooks, like, e.g.,
\cite{weinberg, gmm} that under diffeomorphisms spinor fields transform as
scalars, but we will not comment more on that here.}%
\begin{gather}
\mathfrak{g}_{\mathbf{J}}=1,\mathfrak{g}_{\mathbf{j}_{i}},...,\mathfrak{g}%
_{\mathbf{j}_{1}\mathbf{j}_{2}\mathbf{j}_{3}\mathbf{j}_{4}}=\mathfrak{g}%
_{\mathbf{j}_{1}}\wedge\mathfrak{g}_{\mathbf{j}_{2}}\wedge\mathfrak{g}%
_{\mathbf{j}_{3}}\wedge\mathfrak{g}_{\mathbf{j}_{4}},\nonumber\\
\mathfrak{g}^{\mathbf{J}}=1,\mathfrak{g}^{\mathbf{j}_{1}},...,\mathfrak{g}%
^{\mathbf{j}_{1}\mathbf{j}_{2}\cdots\mathbf{j}_{4}}=\mathfrak{g}%
^{\mathbf{j}_{1}}\wedge\mathfrak{g}^{\mathbf{j}_{2}}\wedge\mathfrak{g}%
^{\mathbf{j}_{3}}\wedge\mathfrak{g}^{\mathbf{j}_{4}}.\label{A3}%
\end{gather}

The scalar product ($\cdot$) and the exterior product extend to all sections
of $\mathcal{C\ell(}M,\mathtt{g})$ and here we distinguish the scalar product
from the operations of left and right contractions. We have for for any
$X,Y\in\sec$ $\mathcal{C\ell(}M,\mathtt{g})$%
\begin{equation}
X\cdot Y=\langle\tilde{X}Y\rangle_{0}=\langle X\tilde{Y}\rangle_{0}=Y\cdot X.
\label{T.55}%
\end{equation}
and for arbitrary multiforms $X,Y,Z\in\sec$ $\mathcal{C\ell(}M,\mathtt{g})$
the left and right contractions $\ $of $X$ and $Y$ are the mappings
$\underset{\boldsymbol{g}}{\lrcorner}:\sec\mathcal{C\ell(}M,\mathtt{g}%
)\times\sec\mathcal{C\ell(}M,\mathtt{g})\rightarrow\sec$ $\mathcal{C\ell
(}M,\mathtt{g})$, $\underset{\boldsymbol{g}}{\llcorner}:\sec\mathcal{C\ell
(}M,\mathtt{g})\times\sec\mathcal{C\ell(}M,\mathtt{g})\rightarrow\sec$
$\mathcal{C\ell(}M,\mathtt{g})$ such that%
\begin{align}
(X\underset{\boldsymbol{g}}{\lrcorner}Y)\cdot Z  &  =Y\underset{\boldsymbol{g}%
}{\cdot}(\tilde{X}\wedge Z),\nonumber\\
(X\underset{\boldsymbol{g}}{\llcorner}Y)\cdot Z  &  =X\underset{\boldsymbol{g}%
}{\cdot}(Z\wedge\tilde{Y}). \label{T49}%
\end{align}

\subsection{Spin-Clifford Bundle and Dirac-Hestenes Spinor Fields}

In \cite{rod2004,moro2004} Dirac-Hestenes spinor fields living in a structure
$(M,\boldsymbol{g})$ are sections of the spin-Clifford bundle $\mathcal{C\ell
}_{\mathrm{Spin}}^{l}\mathcal{(}M,\mathtt{g})=\times P_{\mathrm{Spin}%
_{1,3}^{e}}(M,\boldsymbol{g)}\times_{l}\mathbb{R}_{1,3}^{0}$ and one can show
\ that once we fix a spin coframe a Dirac-Hestenes spinor field $\mathbf{\Psi
}\in\sec\mathcal{C\ell}_{\mathrm{Spin}}^{l}\mathcal{(}M,\mathtt{g})$ has a
representative $\psi\in\sec\mathcal{C\ell}^{0}\mathcal{(}M,\mathtt{g})$, i.e.,
an even section of the Clifford bundle $\mathcal{C\ell(}M,\mathtt{g})$. A
covariant Dirac spinor field $\boldsymbol{\psi}$ used by physicists is a
section of the bundle $P_{\mathrm{Spin}_{1,3}^{e}\times_{D^{1/2.0}\oplus
D^{0,1/2}}}\mathbb{C}^{4}$. Details of the above theory may be found in
\cite{rod2004,moro2004,rodcap,lrw2015}. Below we give a dictionary that one
can use to immediately translate results of the standard matrix formalism in
the language of the Clifford bundle formalism and vice-versa. This dictionary
will help the reader to compare the result we found for the energy-momentum
tensor of the Dirac field in the presence of a gravitational field with other
results on that subject that he may find in the literature.
\begin{align}
\boldsymbol{\gamma}_{\mathbf{a}}\boldsymbol{\psi}  &  \leftrightarrow
\mathfrak{g}_{\mathbf{a}}\psi\mathfrak{g}_{0},\nonumber\\
\mathrm{i}\boldsymbol{\psi}  &  \leftrightarrow\psi\mathfrak{g}_{2}%
\mathfrak{g}_{1},\nonumber\\
\mathrm{i}\boldsymbol{\gamma}_{5}\boldsymbol{\psi}  &  \leftrightarrow
\psi\sigma_{3}=\psi\mathfrak{g}_{3}\mathfrak{g}_{0},\nonumber\\
\boldsymbol{\bar{\psi}}  &  =\boldsymbol{\psi}^{\dagger}\boldsymbol{\gamma
}^{0}\leftrightarrow\tilde{\psi},\nonumber\\
\boldsymbol{\psi}^{\dagger}  &  \leftrightarrow\mathfrak{g}_{0}\tilde{\psi
}\mathfrak{g}_{0},\nonumber\\
\boldsymbol{\psi}^{\ast}  &  \leftrightarrow-\gamma_{2}\psi\gamma_{2}.
\label{mae5}%
\end{align}
where $\boldsymbol{\gamma}_{\mathbf{a}}$, $\mathbf{a=}0,1,2,3$ are Dirac
matrices in standard representation, $\boldsymbol{\gamma}_{5}%
=\boldsymbol{\gamma}_{0}\boldsymbol{\gamma}_{1}\boldsymbol{\gamma}%
_{2}\boldsymbol{\gamma}_{3}$ and $\mathrm{i}=\sqrt{-1}$.

\begin{remark}
Note that $\boldsymbol{\gamma}_{\mathbf{a}},\mathrm{i}\mathbf{1}_{4}$ and the
operations $\overline{}$ and $\dagger$ are for each $x\in M$ mappings
$\mathbb{C}^{4}\rightarrow\mathbb{C}^{4}$. Then they are represented in the
Clifford bundle formalism by extensor fields which maps $\mathcal{C\ell}%
^{0}(M,\eta)$ $\rightarrow\mathcal{C\ell}^{0}(M,\eta)$. Thus, to the operator
$\boldsymbol{\gamma}_{\mathbf{a}}$ there corresponds an extensor field, call
it \underline{$\mathfrak{g}$}$_{\mathbf{a}}:\mathcal{C\ell}^{0}(M,\eta)$
$\rightarrow\mathcal{C\ell}^{0}(M,\eta)$ such that \underline{$\mathfrak{g}$%
}$_{\mathbf{a}}\psi=\mathfrak{g}_{\mathbf{a}}\psi\mathfrak{g}_{\mathbf{0}}$.
\end{remark}

\begin{remark}
Recall that the structure $(M,\boldsymbol{\mathring{g}},\mathring{D}%
,\tau_{\boldsymbol{\mathring{g},}}\uparrow_{e_{0}})$ is \emph{Minkowski
spacetime} when $\mathring{D}$ is the Levi-Civita connection of
$\boldsymbol{\mathring{g}}$, $\tau_{\boldsymbol{\mathring{g},}}=\theta
^{\mathbf{0}}\wedge\theta^{\mathbf{1}}\wedge\theta^{\mathbf{2}}\wedge
\theta^{\mathbf{2}}$ defines a positive orientation for $M$ and $\uparrow
_{e_{0}}$ defines a time orientation (given by the global vector field $e_{0}%
$). Also the structure $(M,\boldsymbol{g},D,\tau_{\boldsymbol{g,}}%
\uparrow_{\mathfrak{e}_{0}})$ is a \emph{Lorentzian spacetime} when $D$ is the
Levi-Civita connection of $\boldsymbol{g}$, $\tau_{\boldsymbol{g,}%
}=\mathfrak{g}^{\mathbf{0}}\mathfrak{g}^{\mathbf{1}}\mathfrak{g}^{\mathbf{2}%
}\mathfrak{g}^{\mathbf{2}}\in\sec%
%TCIMACRO{\tbigwedge ^{r}}%
%BeginExpansion
{\textstyle\bigwedge^{r}}
%EndExpansion
T^{\ast}M\hookrightarrow\sec\mathcal{C\ell(}M,\mathtt{g})$ defines a positive
orientation for $M$ and $\uparrow_{\mathfrak{e}_{0}}$ defines a time
orientation (given by the global vector field $\mathfrak{e}_{0}$).
\end{remark}

\subsection{The Lie Derivative of Clifford and Spinor Fields}

In \cite{lrw2015} we give a geometrical motivated definition for the Lie
derivative of spinor fields in the direction of an arbitrary smooth vector
field $\boldsymbol{\xi}\in\sec TM$ which we called the \emph{spinor Lie
derivative} and denoted $\overset{s}{\pounds _{\boldsymbol{\xi}}}$. Let
$\mathcal{C}\in\sec\mathcal{C\ell(}M,\mathtt{g})$ (Eq.(\ref{a2A})) Then,%
\begin{equation}
\overset{s}{\pounds _{\boldsymbol{\xi}}}\mathcal{C}:\mathcal{=}\mathfrak{d}%
_{_{\boldsymbol{\xi}}}\mathcal{C}+\frac{1}{4}[\mathbf{S}(\boldsymbol{\xi
}),\mathcal{C}].\label{a31}%
\end{equation}
Also, the Lie derivative of a Dirac-Hestenes spinor field $\mathbf{\Psi}%
\in\sec\mathcal{C\ell}_{\mathrm{Spin}}^{l}\mathcal{(}M,\mathtt{g})$ is
\begin{equation}
\overset{s}{\pounds _{\boldsymbol{\xi}}}\mathbf{\Psi}:\mathcal{=}%
\mathfrak{d}_{_{\boldsymbol{\xi}}}\mathbf{\Psi}+\frac{1}{4}\mathbf{S}%
(\boldsymbol{\xi})\mathbf{\Psi.}\label{a32}%
\end{equation}
The spinor Lie derivative of a representative\ $\psi\in\sec\mathcal{C\ell
(}M,\mathtt{g})$ of a Dirac-Hestenes spinor field $\mathbf{\Psi}\in
\sec\mathcal{C\ell}_{\mathrm{Spin}}^{l}\mathcal{(}M,\mathtt{g})$ is denoted
$\overset{(s)}{\pounds _{\boldsymbol{\xi}}}\psi$ and we have
\begin{equation}
\overset{s}{\pounds _{\boldsymbol{\xi}}}\psi:\mathcal{=}\mathfrak{d}%
_{_{\boldsymbol{\xi}}}\psi+\frac{1}{4}\mathbf{S}(\boldsymbol{\xi}%
)\psi.\label{a33}%
\end{equation}

In Eqs.(\ref{a31}),(\ref{a32}) and (\ref{a33}) $\mathfrak{d}%
_{_{\boldsymbol{\xi}}}$ denotes the Pfaff derivative and with $\xi
=\boldsymbol{g}(\boldsymbol{\xi},~)$
\begin{equation}
\mathbf{S}(\boldsymbol{\xi})=L(\boldsymbol{\xi})+d\xi\label{a34}%
\end{equation}
with
\begin{equation}
L(\boldsymbol{\xi})=\frac{1}{2}\left(  c_{\mathbf{akl}}+c_{\mathbf{kal}%
}+c_{\mathbf{lak}}\right)  \xi^{\mathbf{k}}\mathfrak{g}^{\mathbf{a}}%
\wedge\mathfrak{g}^{\mathbf{l}}\label{a35}%
\end{equation}
with $c_{\cdot\mathbf{ab}}^{\mathbf{k\cdot\cdot}}$ the structure coefficients
of the basis $\{\mathfrak{e}_{\mathbf{a}}\}$ of $TM$ dual of the basis
$\{\mathfrak{g}^{\mathbf{a}}\}$ of $%
%TCIMACRO{\tbigwedge \nolimits^{1}}%
%BeginExpansion
{\textstyle\bigwedge\nolimits^{1}}
%EndExpansion
T^{\ast}M$. i.e.,
\begin{equation}
\lbrack\mathfrak{e}_{\mathbf{a}},\mathfrak{e}_{\mathbf{b}}]=c_{\cdot
\mathbf{ab}}^{\mathbf{k\cdot\cdot}}\mathfrak{e}_{\mathbf{k}}%
,~~~~\ d\mathfrak{g}^{\mathbf{k}}=-\frac{1}{2}c_{\cdot\mathbf{ab}%
}^{\mathbf{k\cdot\cdot}}\mathfrak{g}^{\mathbf{a}}\wedge\mathfrak{g}%
^{\mathbf{b}}.\label{a36}%
\end{equation}

\begin{remark}
Our definition of spinor Lie derivative \cite{lrw2015} can be extended also
for some cotensor fields, in particular,$\overset{s}{\pounds _{\boldsymbol{\xi
}}}\boldsymbol{g}=0$ the Lie derivative of the field $\boldsymbol{g}$ is null.
This result is very important for the objective of this paper, where a
variation $\boldsymbol{\delta}\mathfrak{g}^{\mathbf{a}}$ is defined as
\begin{equation}
\boldsymbol{\delta}\mathfrak{g}^{\mathbf{a}}%
=-\overset{s}{\pounds _{\boldsymbol{\xi}}}\mathfrak{g}^{\mathbf{a}}
\label{a37}%
\end{equation}
for appropriate vector fields $\boldsymbol{\xi}$ \emph{(}see below\emph{).}
\end{remark}

\begin{remark}
It is very important to recall that there are several non equivalent
definitions for the Lie derivative of spinor fields. Relevant references are
given in \emph{\cite{lrw2015}}. Here we comment that our spinor Lie derivative
of Clifford and spinor fields is obtained obtaining through the introduction
of a spinor mapping $\overset{s}{\mathrm{h}}$ \ which gives a spinor image of
Clifford and spinor fields between points $x^{\prime}=\mathrm{h}x$ and $x$
\emph{(}where $\mathrm{h}:M\rightarrow M$ is a diffeomorphism generated by an
arbitrary differentiable vector field $\boldsymbol{\xi}$\emph{). It is very
important to emphasize here that if }$\Psi\in\sec\mathcal{C\ell}%
_{\mathrm{Spin}}^{l}\mathcal{(}M,\mathtt{g})=P_{\mathrm{Spin}_{1,3}^{e}%
}(M,\boldsymbol{g})\times_{l}\mathbb{R}_{1,3}^{0}$ then its image
$\overset{s}{\mathrm{h}}\Psi$ is also a section of $\mathcal{C\ell
}_{\mathrm{Spin}}^{l}\mathcal{(}M,\mathtt{g})=P_{\mathrm{Spin}_{1,3}^{e}%
}(M,\boldsymbol{g})\times_{l}\mathbb{R}_{1,3}^{0}$.

\emph{The map }$\overset{s}{\mathrm{h}}$ is different from the pullback map
and for the case of Clifford fields it coincides with the pullback mapping
only when the vector field $\boldsymbol{\xi}$ is a Killing vector field in the
structure $(M,\boldsymbol{g})$.

In particular it is important to emphasize here that
$\overset{s}{\pounds _{\boldsymbol{\xi}}}\boldsymbol{g}=0$, i.e., the Lie
derivative of the metrical field is null,which means that when varying the
gravitational potentials the field $\boldsymbol{g}$ does not change..

We also recall here that is even possible \emph{\cite{daper}} to give a
meaning to a statement found in physical textbooks, e.g.,
\emph{\cite{green,gmm,weinberg}} that spinor fields transform under the
pullback $\mathrm{h}^{\ast}$ mapping \ as scalar functions. Briefly, this is
to be understood in the following way. Let $\boldsymbol{g}$ be a metric field
in $M$ and $\boldsymbol{g}^{\prime}\boldsymbol{=}$\texttt{h}$^{\ast
}\boldsymbol{g}$ the pullback metric under a mapping \texttt{h}$:M\rightarrow
M$. If $\Psi\in\sec\mathcal{C\ell}_{\mathrm{Spin}}^{l}\mathcal{(}%
M,\mathtt{g})=P_{\mathrm{Spin}_{1,3}^{e}}(M,\boldsymbol{g})\times
_{l}\mathbb{R}_{1,3}^{0}$ then $\Psi^{\prime}=\mathtt{h}^{\ast}\Psi\in
\sec\mathcal{C\ell}_{\mathrm{Spin}}^{l}\mathcal{(}M^{\prime},\mathtt{g}%
^{\prime})=P_{\mathrm{Spin}_{1,3}^{e}}(M^{\prime},\boldsymbol{g}^{\prime
})\times_{l}\mathbb{R}_{1,3}^{0}$ is such that $\Psi^{\prime}(x)=\Psi
($\texttt{h}$x)$. This definition, a mathematical legitimate one seems to us
an odd one since in particular $(M,\boldsymbol{g})$ and $(M^{\prime
},\boldsymbol{g}^{\prime})$ are supposed in General Relativity to describe the
same gravitational field even if $M^{\prime}=M$ \emph{(}diffeomorphism
invariance of the theory\emph{)}.
\end{remark}

\section{The Energy-Momentum 1-Forms for the Dirac Field in the Presence of a
Gravitational Field}

Let $\mathcal{F}:\sec%
%TCIMACRO{\tbigwedge \nolimits^{k}}%
%BeginExpansion
{\textstyle\bigwedge\nolimits^{k}}
%EndExpansion
T^{\ast}M\rightarrow\sec%
%TCIMACRO{\tbigwedge \nolimits^{4}}%
%BeginExpansion
{\textstyle\bigwedge\nolimits^{4}}
%EndExpansion
T^{\ast}M$, $X\mapsto\mathcal{F}(X)$ be a differentiable multiform function of
a multiform variable $X$. We recall that the directional
derivative\footnote{If the reader needs details in order to follow the
calculations in this Appendix (which needs many \textquotedblleft tricks of
the trade\textquotedblright\ of the Clifford bundle formalism) he can consult
Chapters 2\ and 7 of \cite{rodcap} and \cite{lrw2015}.} of $\mathcal{F}$ in
the direction of $\mathcal{W}\in\sec%
%TCIMACRO{\tbigwedge }%
%BeginExpansion
{\textstyle\bigwedge}
%EndExpansion
T^{\ast}M$ is denoted$\cdot$ $\mathcal{W\cdot\partial}_{X}\mathcal{F}$ and we have%

\begin{equation}
\mathcal{W\cdot\partial}_{X}\mathcal{F}(X)=\lim_{t\rightarrow0}\frac
{\mathcal{F}(X+t\langle\mathcal{W}\rangle_{X})-\mathcal{F}(X)}{t}. \label{A1}%
\end{equation}

Moreover, the \ multiform derivative $\mathcal{\partial}_{X}\mathcal{F}$ is
defined by\footnote{In \cite{rodcap} we also use the notation $\partial
_{X}\mathcal{F}(X)=\mathcal{F}^{\prime}(X)$.}%

\begin{equation}
\partial_{X}\mathcal{F}(X)=\underset{\mathbf{J}}{%
%TCIMACRO{\dsum }%
%BeginExpansion
{\displaystyle\sum}
%EndExpansion
}\dfrac{1}{\nu(\mathbf{J})!}\mathfrak{g}^{\mathbf{J}}\mathcal{\partial
}_{\mathfrak{g}_{\mathbf{J}}}\mathcal{F}(X)=\underset{J}{%
%TCIMACRO{\dsum }%
%BeginExpansion
{\displaystyle\sum}
%EndExpansion
}\dfrac{1}{\nu(\mathbf{J})!}\mathfrak{g}_{\mathbf{j}}\mathcal{\partial
}_{\mathfrak{g}^{\mathbf{J}}}\mathcal{F}(X),\label{A2}%
\end{equation}
where the symbols $\mathfrak{g}_{\mathbf{J}}$ and $\mathfrak{g}^{\mathbf{J}}$
are defined in Eq.(\ref{A3}) and $\nu(\mathbf{J})=0,1,2,...$ for
$\mathbf{J}=\varnothing,$ $\mathbf{j}_{1},\mathbf{j}_{2},\mathbf{j}%
_{3},\mathbf{j}_{4},...$ where all indices $\mathbf{j}_{1},\mathbf{j}%
_{2},\mathbf{j}_{3},\mathbf{j}_{4}$\ run from $0$ to $3$.

Now, let the Dirac Lagrangian in interaction with the gravitational field be
given by%
\begin{gather}
\mathcal{L}_{D}:\sec%
%TCIMACRO{\tbigwedge \nolimits^{1}}%
%BeginExpansion
{\textstyle\bigwedge\nolimits^{1}}
%EndExpansion
T^{\ast}M\times(\sec%
%TCIMACRO{\tbigwedge \nolimits^{\bigtriangleup}}%
%BeginExpansion
{\textstyle\bigwedge\nolimits^{\bigtriangleup}}
%EndExpansion
T^{\ast}M)^{2}\times(\sec%
%TCIMACRO{\tbigwedge \nolimits^{\bigtriangledown}}%
%BeginExpansion
{\textstyle\bigwedge\nolimits^{\bigtriangledown}}
%EndExpansion
T^{\ast}M)^{2}\rightarrow\sec%
%TCIMACRO{\tbigwedge \nolimits^{4}}%
%BeginExpansion
{\textstyle\bigwedge\nolimits^{4}}
%EndExpansion
T^{\ast}M,\nonumber\\
(\mathfrak{g}^{\mathbf{k}},\psi,\tilde{\psi},\mathfrak{g}^{\mathbf{k}}%
\partial_{\mathfrak{e}_{\mathbf{k}}}\psi,\mathfrak{g}^{\mathbf{k}}%
\partial_{\mathfrak{e}_{\mathbf{k}}}\tilde{\psi})\mapsto\mathcal{L}%
_{D}(\mathfrak{g}^{\mathbf{k}},\psi,\tilde{\psi},\mathfrak{g}^{\mathbf{k}%
}\partial_{\mathfrak{e}_{\mathbf{k}}}\psi,\mathfrak{g}^{\mathbf{k}}%
\partial_{\mathfrak{e}_{\mathbf{k}}}\tilde{\psi})\nonumber\\
\mathcal{L}_{D}=\left\{
\begin{array}
[c]{c}%
(\mathfrak{g}^{\mathbf{k}}\partial_{\mathfrak{e}_{\mathbf{k}}}\tilde{\psi
}\mathfrak{g}^{\mathbf{2}}\mathfrak{g}^{\mathbf{1}})\mathfrak{g}^{\mathbf{0}%
}\cdot\tilde{\psi}-\frac{1}{4}\mathfrak{g}^{\mathbf{k}}\tilde{\psi
}L(\mathfrak{g}_{\mathbf{k}})\mathfrak{g}^{\mathbf{0}}\mathfrak{g}%
^{\mathbf{2}}\mathfrak{g}^{\mathbf{1}}\cdot\tilde{\psi}\\
+\psi\cdot(\mathfrak{g}^{\mathbf{k}}\partial_{\mathfrak{e}_{\mathbf{k}}}%
\psi\mathfrak{g}^{\mathbf{0}}\mathfrak{g}^{\mathbf{2}}\mathfrak{g}%
^{\mathbf{1}})+\frac{1}{4}\psi\cdot(\mathfrak{g}^{\mathbf{k}}L(\mathfrak{g}%
_{\mathbf{k}})\psi\mathfrak{g}^{\mathbf{0}}\mathfrak{g}^{\mathbf{2}%
}\mathfrak{g}^{\mathbf{1}}+m\psi\cdot\tilde{\psi}%
\end{array}
\right\}  \tau_{\boldsymbol{g}} \label{A4}%
\end{gather}
where%
\begin{align}
\sec%
%TCIMACRO{\tbigwedge \nolimits^{\bigtriangleup}}%
%BeginExpansion
{\textstyle\bigwedge\nolimits^{\bigtriangleup}}
%EndExpansion
T^{\ast}M  &  =\sec(%
%TCIMACRO{\tbigwedge \nolimits^{0}}%
%BeginExpansion
{\textstyle\bigwedge\nolimits^{0}}
%EndExpansion
T^{\ast}M+%
%TCIMACRO{\tbigwedge \nolimits^{2}}%
%BeginExpansion
{\textstyle\bigwedge\nolimits^{2}}
%EndExpansion
T^{\ast}M+%
%TCIMACRO{\tbigwedge \nolimits^{4}}%
%BeginExpansion
{\textstyle\bigwedge\nolimits^{4}}
%EndExpansion
T^{\ast}M),\nonumber\\
\sec%
%TCIMACRO{\tbigwedge \nolimits^{\bigtriangledown}}%
%BeginExpansion
{\textstyle\bigwedge\nolimits^{\bigtriangledown}}
%EndExpansion
T^{\ast}M  &  =\sec(%
%TCIMACRO{\tbigwedge \nolimits^{1}}%
%BeginExpansion
{\textstyle\bigwedge\nolimits^{1}}
%EndExpansion
T^{\ast}M+%
%TCIMACRO{\tbigwedge \nolimits^{3}}%
%BeginExpansion
{\textstyle\bigwedge\nolimits^{3}}
%EndExpansion
T^{\ast}) \label{A5}%
\end{align}
and recalling Eq.(\ref{a35}) it is
\begin{equation}
L(\mathfrak{g}_{\mathbf{k}}):=\frac{1}{2}(c_{\mathbf{rks}}+c_{\mathbf{krs}%
}+c_{\mathbf{srk}})\mathfrak{g}^{\mathbf{r}}\wedge\mathfrak{g}^{\mathbf{s}}.
\label{A6}%
\end{equation}

Now, define $\mathcal{L}_{D}=\underset{\boldsymbol{g}}{\star}\mathfrak{L}%
_{D}=\mathfrak{L}_{D}\tau_{\boldsymbol{g}}$ with%
\begin{equation}
\mathfrak{L}_{D}:\sec%
%TCIMACRO{\tbigwedge \nolimits^{1}}%
%BeginExpansion
{\textstyle\bigwedge\nolimits^{1}}
%EndExpansion
T^{\ast}M\times(\sec%
%TCIMACRO{\tbigwedge \nolimits^{\bigtriangleup}}%
%BeginExpansion
{\textstyle\bigwedge\nolimits^{\bigtriangleup}}
%EndExpansion
T^{\ast}M)^{2}\times(\sec%
%TCIMACRO{\tbigwedge \nolimits^{\bigtriangledown}}%
%BeginExpansion
{\textstyle\bigwedge\nolimits^{\bigtriangledown}}
%EndExpansion
T^{\ast}M)^{2}\rightarrow\sec%
%TCIMACRO{\tbigwedge \nolimits^{0}}%
%BeginExpansion
{\textstyle\bigwedge\nolimits^{0}}
%EndExpansion
T^{\ast}M. \label{A8}%
\end{equation}

The variation of $\mathcal{L}_{D}$ induced by the lifting in the spin
structure bundle of the differentiable vector field $\boldsymbol{\xi
=}e_{\mathbf{a}}$ is defined\ by
\begin{equation}
\boldsymbol{\delta}\mathcal{L}_{D}:=%
%TCIMACRO{\tsum \nolimits_{\mathbf{a}}}%
%BeginExpansion
{\textstyle\sum\nolimits_{\mathbf{a}}}
%EndExpansion
\boldsymbol{\delta}\mathfrak{g}^{\mathbf{k}}\wedge\frac{\partial
\mathcal{L}_{D}}{\partial\mathfrak{g}^{\mathbf{k}}}=\boldsymbol{\delta
}\mathfrak{g}^{\mathbf{k}}\wedge\frac{\partial\mathcal{L}_{D}}{\partial
\mathfrak{g}^{\mathbf{k}}}, \label{AA9}%
\end{equation}

On the other hand the variation of $\mathfrak{L}_{D}$ induced by an
arbitrary\ variation $\mathfrak{g}^{\mathbf{k}}\mapsto\mathfrak{g}%
^{\mathbf{k}}+\mathfrak{X}^{\mathbf{k}}$ ($\mathfrak{X}^{\mathbf{k}}\in\sec%
%TCIMACRO{\tbigwedge \nolimits^{1}}%
%BeginExpansion
{\textstyle\bigwedge\nolimits^{1}}
%EndExpansion
T^{\ast}M\hookrightarrow\sec\mathcal{C\ell(}M,\mathtt{g})$)\ is given by the
directional derivative $\mathfrak{X}^{\mathbf{k}}\cdot\partial_{\mathfrak{g}%
_{\mathbf{k}}}$ of $\mathfrak{L}_{D}$, i.e.,
\begin{equation}
\boldsymbol{\delta}\mathfrak{L}_{D}:=\mathfrak{X}^{\mathbf{k}}\cdot
\partial_{\mathfrak{g}^{\mathbf{k}}}\mathfrak{~L}_{D}=\mathfrak{X}%
^{\mathbf{k}}\cdot(\partial_{\mathfrak{g}^{\mathbf{k}}}\mathfrak{L}%
_{D})\label{A10}%
\end{equation}
where we have used the fact that any $F:$ $\sec%
%TCIMACRO{\tbigwedge \nolimits^{\bigtriangleup}}%
%BeginExpansion
{\textstyle\bigwedge\nolimits^{\bigtriangleup}}
%EndExpansion
T^{\ast}M\ni X\mapsto F(X)\in\sec%
%TCIMACRO{\tbigwedge \nolimits^{0}}%
%BeginExpansion
{\textstyle\bigwedge\nolimits^{0}}
%EndExpansion
T^{\ast}M$ it is \cite{rodcap}%
\begin{equation}
X\cdot\partial_{X}~F=X\cdot\left(  \partial_{X}F\right)  \label{A11}%
\end{equation}

So, taking \ $\mathfrak{X}^{\mathbf{k}}=\boldsymbol{\delta}\mathfrak{g}%
^{\mathbf{k}}$ it is%
\begin{equation}
\boldsymbol{\delta}\mathfrak{L}_{D}:=\boldsymbol{\delta}\mathfrak{g}%
^{\mathbf{k}}\cdot\partial_{\mathfrak{g}^{\mathbf{k}}}\mathfrak{~L}%
_{D}=\boldsymbol{\delta}\mathfrak{g}^{\mathbf{k}}\cdot(\partial_{\mathfrak{g}%
^{\mathbf{k}}}\mathfrak{L}_{D})=\boldsymbol{\delta}\mathfrak{g}^{\mathbf{k}%
}\cdot\langle\partial_{\mathfrak{g}^{\mathbf{k}}}\mathfrak{L}_{D}\rangle
_{1}\label{A11a}%
\end{equation}
and since as it is easy to show $\boldsymbol{\delta}\mathfrak{\tau
}_{\boldsymbol{g}}=-$\ $\overset{s}{\pounds }\mathfrak{\tau}_{\boldsymbol{g}%
}=0$ we can write%
\begin{equation}
\boldsymbol{\delta}\mathcal{L}_{D}=\boldsymbol{\delta}\left(  \mathfrak{L}%
_{D}\tau_{\boldsymbol{g}}\right)  =\left(  \boldsymbol{\delta}\mathfrak{L}%
_{D}\right)  \tau_{\boldsymbol{g}}+\mathfrak{L}_{D}\boldsymbol{\delta}%
\tau_{\boldsymbol{g}}=\left(  \boldsymbol{\delta}\mathfrak{L}_{D}\right)
\tau_{\boldsymbol{g}}=\underset{\boldsymbol{g}}{\star}\boldsymbol{\delta
}\mathfrak{L}_{D}.\label{A12}%
\end{equation}
From Eq.(\ref{A12}) we get
\begin{equation}
\boldsymbol{\delta}\mathcal{L}_{D}=\boldsymbol{\delta}\mathfrak{g}%
^{\mathbf{k}}\wedge\frac{\partial\mathcal{L}_{D}}{\partial\mathfrak{g}%
^{\mathbf{k}}}=\left(  \boldsymbol{\delta}\mathfrak{g}^{\mathbf{k}}%
\cdot(\partial_{\mathfrak{g}_{\mathbf{k}}}\mathfrak{L}_{D})\right)
\tau_{\boldsymbol{g}}=\boldsymbol{\delta}\mathfrak{g}^{\mathbf{k}}%
\wedge\underset{\boldsymbol{g}}{\star}\langle\partial_{\mathfrak{g}%
_{\mathbf{k}}}\mathfrak{L}_{D}\rangle_{1}\label{A13}%
\end{equation}
and since by definition the $1$-forms of energy-momentum
$\overset{D}{\mathcal{T}}_{\mathbf{k}}=\overset{D}{T}_{\mathbf{km}%
}\mathfrak{g}^{\mathbf{m}}$ of the Dirac field in the presence of the
gravitational field are defined by
\begin{equation}
\underset{\boldsymbol{g}}{\star}\overset{D}{\mathcal{T}}_{\mathbf{k}}%
=\frac{\partial\mathcal{L}_{D}}{\partial\mathfrak{g}^{\mathbf{k}}}.\label{A14}%
\end{equation}
We get \ using Eq.(\ref{A13}) the notable relation
\begin{equation}
\overset{D}{\mathcal{T}}_{\mathbf{k}}=\langle\partial_{\mathfrak{g}%
_{\mathbf{k}}}\mathfrak{L}_{D}\rangle_{1}\label{A15}%
\end{equation}
that%

\begin{equation}
\overset{D}{\mathcal{T}}_{\mathbf{k}}=\langle\mathbf{D}_{\mathfrak{e}%
_{\mathbf{k}}}\tilde{\psi}\mathfrak{g}^{\mathbf{2}}\mathfrak{g}^{\mathbf{1}%
}\mathfrak{g}^{\mathbf{0}}\tilde{\psi}+\psi\mathbf{D}_{\mathfrak{e}%
_{\mathbf{k}}}\psi\mathfrak{g}^{\mathbf{0}}\mathfrak{g}^{\mathbf{2}%
}\mathfrak{g}^{\mathbf{1}}\rangle_{1} \label{A16N}%
\end{equation}
where
\begin{equation}
\mathbf{D}_{\mathfrak{e}_{\mathbf{k}}}\psi:=\partial_{\mathfrak{e}%
_{\mathbf{k}}}\psi+\frac{1}{4}L(\mathfrak{g}_{k})\psi. \label{A16a}%
\end{equation}

Moreover, recalling as observed in Section 2 that by Einstein equation (for
the system gravitational plus Dirac field) is $\underset{\boldsymbol{g}%
}{\star}G_{\mathbf{k}}=-\underset{\boldsymbol{g}}{\star}%
\overset{D}{\mathcal{T}}_{\mathbf{k}}$ and $G_{\mathbf{k}}=G_{\mathbf{km}%
}\mathfrak{g}^{\mathbf{m}}$ with $G_{\mathbf{km}}=G_{\mathbf{mk}}$ it follows
that $\overset{D}{\mathcal{T}}_{\mathbf{km}}=\overset{D}{\mathcal{T}%
}_{\mathbf{mk}}$, i.e., $\overset{D}{\mathcal{T}}_{\mathbf{k}}\cdot
\mathfrak{g}_{\mathbf{m}}=\overset{D}{\mathcal{T}}_{\mathbf{m}}\cdot
\mathfrak{g}_{\mathbf{k}}$ . Observe that since for any, $A,B\in
\sec\mathcal{C\ell}(M,\mathtt{g})$ it is $\langle AB\rangle_{r}=(-1)^{r\frac
{(r-1)}{2}}\langle\tilde{B}\tilde{A}\rangle_{r}$ we can write%

\begin{align}
\langle\mathfrak{g}_{\mathbf{m}}\psi\mathbf{D}_{\mathfrak{e}_{\mathbf{k}}}%
\psi\mathfrak{g}^{\mathbf{2}}\mathfrak{g}^{\mathbf{1}}\mathfrak{g}%
^{\mathbf{0}}\rangle_{0}  &  =\langle(\langle\mathfrak{g}_{\mathbf{m}}%
\psi\rangle_{1}+\langle\mathfrak{g}_{\mathbf{m}}\psi\rangle_{3})\mathbf{D}%
_{\mathfrak{e}_{\mathbf{k}}}\psi\mathfrak{g}^{\mathbf{2}}\mathfrak{g}%
^{\mathbf{1}}\mathfrak{g}^{\mathbf{0}}\rangle_{0}\nonumber\\
&  =\langle(\langle\tilde{\psi}\mathfrak{g}_{\mathbf{m}}\rangle_{1}%
+\langle\tilde{\psi}\mathfrak{g}_{\mathbf{m}}\rangle_{3})\mathbf{D}%
_{\mathfrak{e}_{\mathbf{k}}}\psi\mathfrak{g}^{\mathbf{2}}\mathfrak{g}%
^{\mathbf{1}}\mathfrak{g}^{\mathbf{0}}\rangle_{0}=\langle\tilde{\psi
}\mathfrak{g}_{\mathbf{m}}\mathbf{D}_{\mathfrak{e}_{\mathbf{k}}}%
\psi\mathfrak{g}^{\mathbf{2}}\mathfrak{g}^{\mathbf{1}}\mathfrak{g}%
^{\mathbf{0}}\rangle\label{a16n2}%
\end{align}
Also,%

\begin{align}
\langle\mathbf{D}_{\mathfrak{e}_{\mathbf{k}}}\tilde{\psi}\mathfrak{g}%
^{\mathbf{2}}\mathfrak{g}^{\mathbf{1}}\mathfrak{g}^{\mathbf{0}}\tilde{\psi
}\mathfrak{g}_{\mathbf{m}}\rangle_{0}  &  =\langle\mathbf{D}_{\mathfrak{e}%
_{\mathbf{k}}}\tilde{\psi}\mathfrak{g}^{\mathbf{2}}\mathfrak{g}^{\mathbf{1}%
}\mathfrak{g}^{\mathbf{0}}\mathfrak{g}_{\mathbf{m}}\psi\rangle_{0}\nonumber\\
&  =-\langle\mathbf{D}_{\mathfrak{e}_{\mathbf{k}}}\tilde{\psi}\mathfrak{g}%
_{\mathbf{m}}\psi\mathfrak{g}^{\mathbf{2}}\mathfrak{g}^{\mathbf{1}%
}\mathfrak{g}^{\mathbf{0}}\rangle_{0} \label{A16N1}%
\end{align}

So, using the above results we get%
\begin{equation}
\overset{D}{\mathcal{T}}_{\mathbf{mk}}=\frac{1}{2}\left(
\overset{D}{\mathcal{T}}_{\mathbf{k}}\cdot\mathfrak{g}_{\mathbf{m}%
}+\overset{D}{\mathcal{T}}_{\mathbf{m}}\cdot\mathfrak{g}_{\mathbf{k}}\right)
=\frac{1}{2}\langle\tilde{\psi}\mathfrak{g}_{(\mathbf{m}}\mathbf{D}%
_{\mathfrak{e}_{\mathbf{k)}}}\psi\mathfrak{g}^{\mathbf{2}}\mathfrak{g}%
^{\mathbf{1}}\mathfrak{g}^{\mathbf{0}}-\mathbf{D}_{(\mathfrak{e}_{\mathbf{k}}%
}\tilde{\psi}\mathfrak{g}_{\mathbf{k)}}\mathfrak{g}^{\mathbf{2}}%
\mathfrak{g}^{\mathbf{1}}\mathfrak{g}^{\mathbf{0}}\psi\rangle_{0}. \label{A17}%
\end{equation}

\begin{remark}
Using the dictionary \emph{(Appendix A) }between the standard matrix formalism
used by \ physicists for dealing with \emph{(}covariant\emph{)} Dirac spinor
fields and the formalism of this paper where these objects are represented
(once we fix a spin frame\emph{)} by an even section $\psi$ of the Clifford
bundle $\mathcal{C\ell(}M,\mathtt{g})$ we immediately verify that
\emph{Eq.(\ref{A17})} coincides, e.g. with the result reported in
\cite{hannibal}.
\end{remark}

\section{Energy-Momentum of the Gravitational Field for the Schwarzschild
Field}

The Schwarzschild solution  $\boldsymbol{g}$ (of Einstein equation) for a star
of mass $m$ with radius $\mathbf{R}$ greater than the Schwarzschild radius can
be written in polar coordinates covering the region of interest as%

\begin{gather}
\boldsymbol{g}=g_{\mu\nu}dx^{\mu}\otimes dx^{\nu},\nonumber\\
\boldsymbol{g}=\left(  1-\frac{2m}{r}\right)  dt\otimes dt-\left(  1-\frac
{2m}{r}\right)  ^{-1}dr\otimes dr-r^{2}d\theta\otimes d\theta-\left(
r^{2}\sin^{2}\theta\right)  d\varphi\otimes d\varphi.\label{c1}%
\end{gather}

Here according to the theory presented above the gravitational potentials
$\mathfrak{g}^{\mathbf{a}}$, $\mathbf{a}=0,1,2,3$ are:%

\begin{equation}
\mathfrak{g}^{0}=\left(  1-\frac{2m}{r}\right)  ^{\frac{1}{2}}%
dt;~~\mathfrak{g}^{1}=\left(  1-\frac{2m}{r}\right)  ^{-\frac{1}{2}%
}dr;~~\mathfrak{g}^{2}=rd\theta;~~\mathfrak{g}^{3}=r\sin\theta d\varphi
.\label{c2}%
\end{equation}

Using the nice formula Eq.(\ref{2.14}) we will evaluate the energy-momentum
$1$-forms of the Schwarzschild field.

Since the scalar curvature $R=0$ outside the star we have%

\begin{align}
\mathbf{t}^{0}  &  =\partial\cdot\partial~\mathfrak{g}^{0}+d\delta
\mathfrak{g}^{0}=\frac{M^{2}}{\sqrt{1-\frac{2M}{r}}r^{4}}dx^{0}=\frac{M^{2}%
}{\left(  1-\frac{2M}{r}\right)  r^{4}}\mathfrak{g}^{0},\nonumber\\
\mathbf{t}^{1}  &  =\partial\cdot\partial~\mathfrak{g}^{1}+d\delta
\mathfrak{g}^{1}=0,\nonumber\\
\mathbf{t}^{2}  &  =\partial\cdot\partial~\mathfrak{g}^{2}+d\delta
\mathfrak{g}^{2}=\frac{\cot(\theta)}{r^{2}}\left(  1-\frac{2M}{r}\right)
^{\frac{1}{2}}\mathfrak{g}^{1}-\frac{2M}{r^{3}}\mathfrak{g}^{2},\nonumber\\
\mathbf{t}^{3}  &  =\partial\cdot\partial~\mathfrak{g}^{3}+d\delta
\mathfrak{g}^{3}=\frac{-M+r+M\cos(2\theta)}{r^{3}}\csc^{2}(\theta
)\mathfrak{g}^{3}. \label{c4}%
\end{align}
With a simple calculus we see that
\begin{equation}
0=\mathfrak{g}^{2}\cdot\mathbf{t}^{1}=\mathfrak{t}^{21}\neq\mathfrak{t}%
^{12}=\mathfrak{g}^{1}\cdot\mathbf{t}^{2}=-\frac{\cot(\theta)}{r^{2}}\left(
1-\frac{2M}{r}\right)  ^{\frac{1}{2}}. \label{c5}%
\end{equation}

Now it is easy to evaluate the energy of the Schwarzschild gravitational field
outside the star\footnote{We denote this region by $\Omega.$}. We have taking
into account \ the convention used for the definition of the energy-momentum
$3$-forms of the fields (Eq.(\ref{2.v3})) and the equation of motion for the
gravitational potentials (Eq.(\ref{10.12})) that we must define the energy of
the field%

\begin{equation}
E:=-\int_{\Omega}\mathbf{t}^{0}\cdot\mathfrak{g}^{0}dV=-\int_{0}^{2\pi}%
\int_{0}^{\pi}\int_{\mathbf{R}}^{\infty}\frac{M^{2}}{r^{2}\left(  1-\frac
{2M}{r}\right)  ^{\frac{3}{2}}}\sin\theta~drd\theta d\varphi=4\pi M\left[
1-\frac{1}{\left(  1-\frac{2M}{\mathbf{R}}\right)  ^{\frac{1}{2}}}\right]
.\label{c7}%
\end{equation}
For sun's like stars $\frac{2M}{\mathbf{R}}\approx5.10^{6}$. For such cases we
have to first order in $\frac{2M}{\mathbf{R}}$ that
\begin{equation}
E=-4\pi\frac{M^{2}}{\mathbf{R}}.\label{c.8}%
\end{equation}

\begin{remark}
From \emph{Eq.(\ref{c.8})} \ we see that the energy of the gravitational field
in the exterior of the star is negative. The idea that the energy of the
gravitational field is negative is an old one. It appears , e.g., in the Tryon
paper \cite{tryon} which suggested that the universe appears from nothing
through a vacuum fluctuation and also it is essential for the inflationary
cosmology \cite{guth}. And indeed if we supposed that the spatial part of our
universe is closed, e.g., is \footnote{Recall that $S^{3}$ is a parallelizable
manifod for which a small modification of our theory works as well.}$S^{3}$ we
immediately get from \emph{Eq.(\ref{10.16})} and Stokes theorem that since
$\partial S^{3}=\varnothing$ it is
\begin{equation}
\int_{S^{3}}\underset{\boldsymbol{g}}{\star}(\overset{m}{\mathcal{T}}%
_{0}+\mathfrak{t}_{0})=-\int_{S^{3}}d\underset{\boldsymbol{g}}{\star
}\mathcal{F}_{0}=-\int_{\partial S^{3}}\underset{\boldsymbol{g}}{\star
}\mathcal{F}_{0}=0, \label{c.9}%
\end{equation}
which shows that the total gravitational energy of this universe is null,
i.e., the energy-momentum of the gravitational field is negative.
\end{remark}

And, all the momentum components $P^{i}:=-\int_{\Omega}\mathfrak{t}^{0i}dV=0$
are trivially zero.
\end{document}